\documentclass[namedreferences]{SolarPhysics}

\usepackage[hyperref,optionalrh,solaromanenum]{spr-sola-addons}
\usepackage{graphicx}        % For eps figures, newer & more powerfull
\usepackage{color}           % For color text: \color command
\usepackage{breakurl}        % For breaking URLs easily trough lines
\newcommand{\etal}{{\it et al.}}

\usepackage{epstopdf}

\begin{document}
\begin{article}
\begin{opening}

\title{The Relation between Magnetic Fields and X-ray Emission for Solar Microflares and Active Regions}
%\\ {\it Solar Physics}}

%\author[addressref={aff},corref,email={kirichenko@lebedev.ru}]{\inits{A.S.}\fnm{A.S.}~\lnm{Kirichenko}}%\sep

\author[addressref={aff},corref,email={kirichenko@lebedev.ru}]{\inits{A.S.}\fnm{A.S.}\lnm{Kirichenko}\orcid{0000-0002-7303-4098}}

\author[addressref=aff,email={bogachev@lebedev.ru}]{\inits{S.A}\fnm{S.A}~\lnm{Bogachev}}%\sep

%\author{\inits{}\fnm{}~\lnm{}\orcid{}}
%\author{P.~\surname{Author-a}$^{1}$\sep
%        E.~\surname{Author-b}$^{1}$\sep
%        M.~\surname{Author-c}$^{2}$      
%       }

%   \institute{$^{1}$ First affiliation
%                     email: \url{e.mail-a} email: \url{e.mail-b}\\ 
%              $^{2}$ Second affiliation
%                     email: \url{e.mail-c} \\
%             }
\address[id=aff]{P.N. Lebedev physical institute}

\runningauthor{A.S.Kirichenko, S.A.Bogachev}
\runningtitle{Relation between magnetic field and X-ray emission}

%%%%%%%%%%%%%%%%%%%%%%%%%%%%%%%%%%%%%%%%%%%%%%%%%%%
%% Authors Names
%
% \author[addressref={},corref,email={}]{\inits{}\fnm{}\lnm{}\orcid{}}

%%%%%%%%%%%%%%%%%%%%%%%%%%%%%%%%%%%%%%%%%%%%%%%%%%%
%% Affilations 
%% id shold be the same with \author addressref value.
%\address[id={}]{}

%\runningauthor{A.S.~\surname{Kirichenko} and S.A.~\surname{Bogachev}}
\runningtitle{The relation between magnetic fields and X-ray emission for solar microflares and active regions}
\begin{abstract}
We present the result of comparison between magnetic field parameters and the intensity of X-ray emission for solar microflares with \textit{Geosynchronous Operational Environmental Satellites} (GOES) classes from A0.02 to B5.1. For our study, we used the monochromatic \textit{MgXII Imaging Spectroheliometer} (MISH), \textit{Full-disk EUV Telescope} (FET) and \textit{Solar PHotometer in X-rays} (SphinX) instruments onboard the \textit{Complex Orbital Observations Near-Earth of Activity of the Sun-Photon} CORONAS-\textit{Photon} spacecraft because of their high sensitivity in soft X-rays. The peak flare flux (PFF) for solar microflares was found to depend on the strength of the magnetic field and total unsigned magnetic flux as a power-law function. In the spectral range 2.8\,--\,36.6~\AA\ which shows very little increase related to microflares the power-law index of the relation between the X-ray flux and magnetic flux for active regions  is 1.48 $\pm$ 0.86, which is close to the value obtained previously by Pevtsov \etal (\textit{Astrophys. J.} \textbf{598}, 1387, 2003) for different types of solar and stellar objects. In the spectral range 1\,--\,8~\AA\ the power-law indices for PFF($B$) and PFF($\Phi$) for microflares are 3.87 $\pm$ 2.16 and 3 $\pm$ 1.6 respectively. We also make suggestions on the heating mechanisms in active regions and microflares under the assumption of loops with constant pressure and heating using the Rosner-Tucker-Vaiana (RTV) scaling laws.
\end{abstract}
\end{opening}

\section{Introduction}
     \label{S-Introduction} 
The mechanism of plasma heating in the solar corona is one of the fundamental problems of modern solar physics. This process is still not known about in detail, but it is obvious that the magnetic field plays a key role in the corresponding energy storage and release. Since the energy for plasma heating is drawn from the magnetic field, there should be a relationship between the parameters of the magnetic field (for example, its strength or flux) and the radiative characteristics of plasma (luminosity, temperature and emission measure). The observations demonstrate that such a relationship really exists. \citet{Golub1980} analyzed active regions and X-ray bright points (XBP) using \textit{Skylab} data. They obtained a power-law relationship with index 1.5 between thermal energy $E_{\rm {th}}$ and the magnetic flux $ \Phi $ in the active regions and in XBPs. It was shown that the plasma pressure also depends on the magnetic field strength $B$ as a power-law function with index 1.6. A similar problem was investigated by \citet{Yashiro2001}. They used \textit{Soft X-ray Telescope} (SXT) data for 31 active regions and obtained indices 1.33 and 0.78 for $E_{\rm th}(\Phi)$ and $P(B)$ respectively, where $p$ is the pressure. They made some suggestions on the coronal heating mechanisms using these results. Coronal heating can be described with the following relationship:
\begin{equation}\label{1}
F\propto B^{\alpha}l^{\beta}
\end{equation}
where $ F $ is the heating flux, $ B $ is the magnetic field strength, $ l $ is the loop length and $\alpha$ and $\beta$ are coefficients which depend on the heating mechanisms. For example, for the nanoflare heating scenario $\alpha=2$ and for Alfv\'en wave heating $\alpha=1$ \citep{Galsgaard1996, Fisher1998, Yashiro2001}. \citet{Yashiro2001} used the Rosner-Tucker-Vaiana (RTV) scaling laws \citep{Rosner1978} for the observed gas pressure $ p $, volumetric heating rate $E_{\rm H}$, loop temperature $ T $ and length $ l $:
\begin{equation}\label{2}
T\approx1.4\times10^{3}(pl)^{1/3}
\end{equation}
\begin{equation}\label{3}
E_{H}\approx10^{5}p^{7/6}l^{-5/6}.
\end{equation}
Taking into account Equations (\ref{1}), (\ref{2}) and (\ref{3}) they obtained the following relationship:
\begin{equation}\label{4}
p\propto B^{6\alpha/7}l^{(6\beta-1)/7}.
\end{equation}
They demonstrated that the correlation between $B$ and $l$ was weak, therefore they considered the relation $p(B)$ independently from the $l$. From the experimental relation $p \approx B^{0.78\pm0.23}$ and Equation (\ref{4}) they found $\alpha=0.91\pm0.27$. So, their result is more consistent with the Alfv\'en wave heating model. Of course, the investigation of heating mechanisms is a more complex problem than a simple comparison of the power-law index with fixed values of 1 or 2. As an example, one can see \citet{Mandrini2000}, where the dependence of the heating rate on \textit{B} and other loop parameters was analyzed for over 20 different heating mechanisms. \citet{Fisher1998} analyzed 333 active regions using SXT data and derived a power-law index 1.19 for the relationship between the soft X-ray luminosity of plasma $L_{x}$ and the magnetic flux of active regions. They noted that their result is consistent with the minimum current corona (MCC) \citep{Longcope1996} and Alfv\'en wave models, although Alfv\'en heating is probably insufficient.\citet{Wolfson2000} obtained the index 1.86 for the $L_{x}(\Phi)$ relationship using SXT data averages along longitude for the whole disk. Comparing their results with theoretical models they refer to the model of \citet{Roald2000}, that links coronal heating with reconnection in the chromospheric network. A similar analysis was performed by \citet{Benevolenskaya2002} for SXT and \textit{Extreme ultraviolet Imaging Telescope} (EIT) data collected during a 10-year period of observations. They derived that the slope of $L_{x}(B)$ depends on the phase of the solar cycle and is equal to 1.5 in maximum and 2.2 in minimum. \citet{Fludra2008} suggest that the difference in slopes found by \citet{Benevolenskaya2002} between the minimum and maximum of solar cycle may be explained by the averaging of too large areas of the solar disk, which contain a mixture of quiet sun and active regions. \citet{Pevtsov2003} considered quiet Sun, X-ray bright points, active regions and different classes of stars and found that the relationship $L_{x}(\Phi)$ for all those classes is a power law function with an index of 1.13 in the energy range 2.8-36.6~\AA. This result is consistent with \citet{Fisher1998} data. \citet{Fludra2008} demonstrated that the power-law index for $L(\Phi)$ depends on the spectral range. They analyzed 48 active regions using a base of CDS/SOHO data and calculated the $L(\Phi)$ indexes for different spectral lines. They found that the slope is greater than 1 for coronal lines and less than 1 for transition region lines, which follows from a different dependence of the line intensity on pressure. Under the assumption of hydrostatic loops they placed a constraint on the coronal heating models, obtaining the relation $E_{\rm H}\propto B^{\gamma}$ with $0.6<\gamma<1.1$, where $E_{\rm H}$ is a volumetric heating rate. \citet{Su2007} analyzed 31 two-ribbon flares of M-class and higher, accompanied by coronal mass ejections (CMEs), considering the correlations between peak flare flux (PFF) and magnetic field strength PFF$(B)$ and between PFF and the magnetic flux PFF$(\Phi)$. The indices for PFF$(B)$ and PFF$(\Phi)$ were found to be 0.93 and 1.1 respectively. They didn't provide such information for weak events of A-C class and lower, because of the lack of corresponding data. In general, we can see that despite the variety of works devoted to the correlation between thermal and magnetic properties of different structures, there is a lack of information for flares and, especially, for weak events, such as microflares. In this work we try to fill this gap and especially focus our attention on microflares. We also try to make some suggestions on the heating mechanisms in microflares using the obtained relations.
\section{Data Selection and Processing}
     \label{S-Data} 
\begin{table}
	\caption{The relations between thermal and magnetic parameters for different solar structures. $P$ is pressure, $U_{\rm th}$ is thermal energy, $I$ is intensity, $L_{x}$ is X-ray emission, and $F$ is X-ray flux.}
   \label{table1}	
	\begin{tabular}{lcccc}
	\hline
	Authors & $f$($ B $)& $f$($ \Phi $)  & Objects & Wavelengths,~\AA \\
	\hline
	\citet{Golub1980} & $P\propto B^{1.6}$ & $U_{\rm th}\propto \Phi^{1.6}$ & AR, XBP & 2-60  \\
	\citet{Fisher1998}  &  & $L_{x}\propto \Phi^{1.19}$ & AR & 1\,--\,300 \\
	\citet{Wolfson2000}  & $F\propto B^{1.86}$ &  & DA & 2.8\,--\,36.6 \\
	\citet{Yashiro2001}    & $P\propto B^{0.78}$ & $U_{\rm th}\propto \Phi^{1.33}$ & AR & 2.8\,--\,36.6  \\
	\citet{Benevolenskaya2002}  & $I\propto B^{2.2}$ &  & DA & 6\,--\,13 \\
	\citet{Pevtsov2003}  &  & $L_{x}\propto \Phi^{1.13}$ & AR, QS, & 2.8\,--\,36.6 \\
	   &  &  & XBP, DA &  \\
	\citet{Fludra2008}   &  & $I\propto \Phi^{0.76}$  & AR & 629.7 \\ %(O V)
	   &  & $I\propto \Phi^{0.88}$ & AR & 584.3 \\ %(He I)
	   &  & $I\propto \Phi^{0.74}$ & AR & 368.06 \\ %(Mg IX)
	   &  & $I\propto \Phi^{1.32}$ & AR & 360.76 \\	 %(Fe XVI)
	\citet{Su2007}  & $F\propto B^{0.93}$ & $F\propto \Phi^{1.1}$ & Flares & 1\,--\,8  \\
	\hline
\multicolumn{5}{l}{AR - active region, XBP - X-ray bright point, QS - quiet Sun, DA - disk average}
	\end{tabular}
\end{table}
\subsection{Data Selection}
For our analysis we used data from 4 space instruments: the  \textit{Michelson Doppler Imager} (MDI) onboard \textit{Solar and Heliospheric Observatory} (SOHO) and 3 instruments from the CORONAS-\textit{Photon} spacecraft \,--\, the \textit{Solar PHotometer in X-rays} (SphinX) \citep{Sylwester2008,Gryciuk2017}, \textit{MgXII Imaging Spectroheliometer} (MISH) \citep{Kirichenko2017}, and \textit{Full-disk EUV Telescope} (FET) \citep{Kuzin2009}. The MDI data were used to calculate the strength of magnetic field and the magnetic flux in the flare regions. To clarify the contour of the flaring region and to obtain the lengths of flaring loops we used FET images (Fe{\sc xxiii} 132~\AA). The data from SphinX and MISH were used to calculate the flare luminosity in different spectral ranges. The period of observations from April to July 2009 was selected due to the low level of solar activity, which was favorable for registration of weak flares. For this period, using the SphinX events catalog, we selected the flares of B-class and lower with the time gap between flare maximum and the nearest MISH image of less than 1 minute. As a result, we selected 163 microflares for further analysis.
\subsection{Processing of SphinX Data}

SphinX is an X-ray spectrophotometer onboard the CORONAS-Photon spacecraft that registered X-ray emission in the energy range 0.5-15 keV. The significant advantage of SphinX was its high sensitivity, which allowed it to register events 1-2 orders of magnitude smaller that the A1.0 level. SphinX operated in two main modes: time stamping and spectral. In the first regime the arrival time and the energy for each photon were recorded. In the second mode, the SphinX accumulated photon spectra over some time interval. The exposure time in this mode varies from one observation to another. 

\begin{figure}    %%%%%%%%%%%%%%%%%% FIGURE 1 
   \centerline{\includegraphics[width=1\textwidth,clip=]{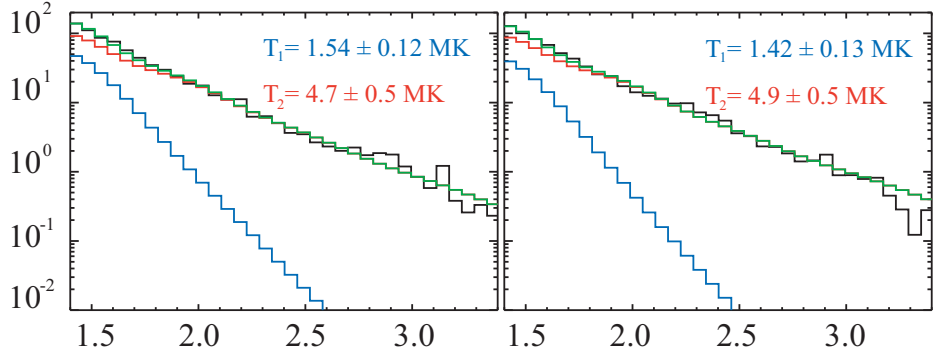}
              }
              \caption{Spectra at the maximum of microflares registered on 8 June 2009 with maximum at 17:54 (left plot) and on 19 April with maximum at 02:08 (right plot). Blue and red spectra are the cold and hot components of the spectral fitting by the 2T model. Green spectra are the sum of both components. 
                      }
   \label{spectrum_2t}
   \end{figure}

\begin{figure}    %%%%%%%%%%%%%%%%%% FIGURE 2
   \centerline{\includegraphics[width=1\textwidth,clip=]{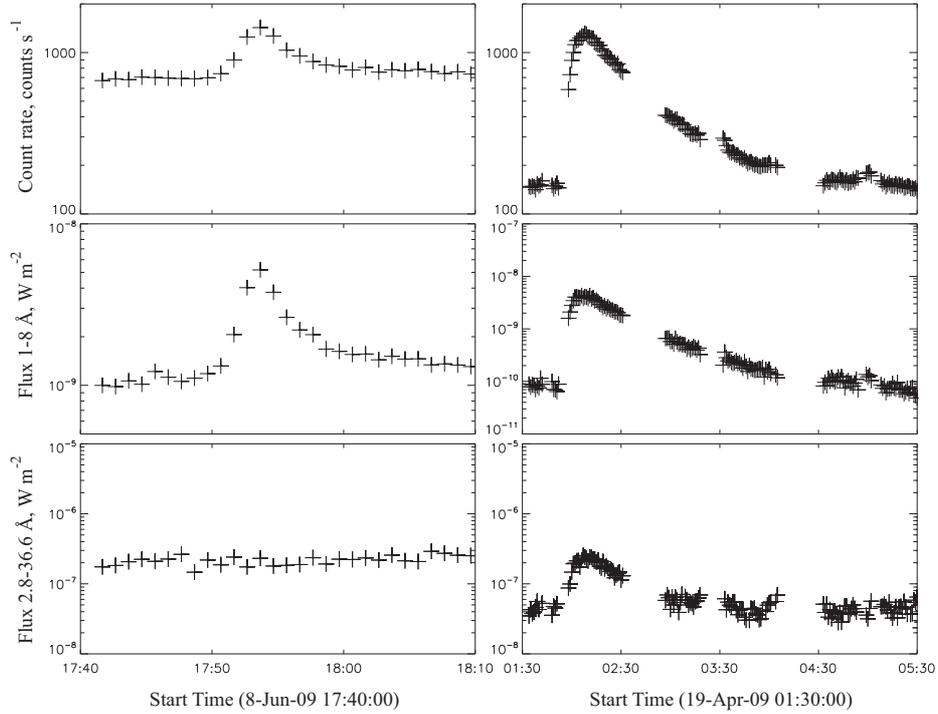}
              }
              \caption{Time profiles for two microflares registered on 8 June 2009 with maximum at 17:54 (left plots) and on 19 April with maximum at 02:08 (right plots): top panels \,--\, integrated SphinX count rate; middle panels \,--\, X-ray flux in the range 1\,--\,8~\AA; bottom panels \,--\, X-ray flux in the range 2.8\,--\,36.6~\AA.
                      }
   \label{profile}
   \end{figure}

\begin{figure}    %%%%%%%%%%%%%%%%%% FIGURE 3 
   \centerline{\includegraphics[width=1\textwidth,clip=]{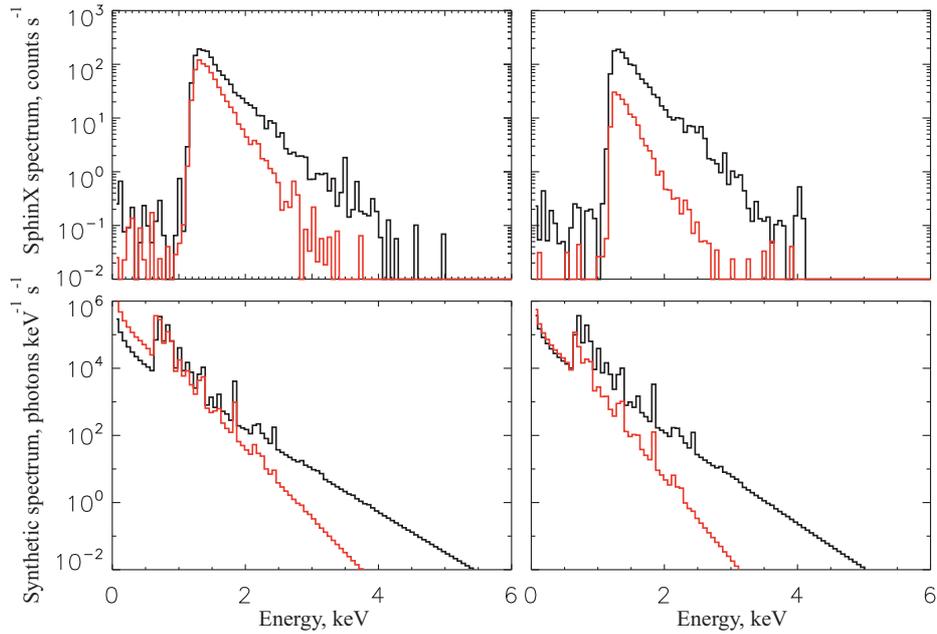}
              }
              \caption{Spectra at the maximum (black color) and preflare background spectra (red color) of microflares registered on 8 June 2009 with maximum at 17:54 (left plots) and on 19 April with maximum at 02:08 (right plots). Top panel: SphinX data in counts per second. Bottom panel: synthetic spectra.
                      }
   \label{spectrum_1t}
   \end{figure}

We used SphinX data to determine the temperature and the emission measure of plasma in the active region just before the flare and at the moment of flare maximum. The preflare spectra were fitted by one thermal component. The spectra at the moment of flare maximum were considered in the two-temperature approximation (an example of such a fit is shown in Figure~\ref{spectrum_2t}). The hot component in this case was considered as a flaring one. For more accurate spectral analysis we increased the statistics by integration of spectra over 30 second periods. Thus, the temporal resolution and the accuracy for the flare maximum time was 30 seconds. Then, we constructed the synthetic spectra with calculated temperature and emission measure by applying the CHIANTI atomic database \citep{Dere1997,Landi2013} and used them to integrate the X-ray emission of the flare in two spectral ranges:  2.8\,--\,36.6~\AA\ and 1\,--\,8~\AA. We chose the range 2.8\,--\,36.6~\AA\ to compare our results with some previous papers mentioned in the introduction. The range 1\,--\,8~\AA\ is a \textit{Geosynchronous Operational Environmental Satellites} (GOES) standard to study solar flares and to obtain their X-ray class. The second reason to study this spectral range in detail is that the emission in 1\,--\,8~\AA\ is usually very sensitive to high temperature plasma in flares. Due to this fact, the flare profiles in 1\,--\,8~\AA\ are usually much more pronounced than ones in lower energy ranges. We found only several flares clearly seen in the range 2.8\,--\,36.6~\AA, while all the flares studied were well detected in the range 1\,--\,8~\AA. To be more clear in how a typical microflare looks in different spectral ranges we present two examples in Figure~\ref{profile}. The left and right plots correspond to different events. The top panels contain the count rate data for both events obtained by integrating all the SphinX channels. Both flares are well distinguished in such a presentation. The time profiles of flares in the ranges 1\,--\,8~\AA\ and 2.8\,--\,36.6~\AA\ obtained in the one-temperature (1T) approximation are plotted in the middle and bottom panels respectively. The first microflare is not seen in the range 2.8\,--\,36.6~\AA\ which is typical for the 1T approximation as mentioned above. Figure~\ref{spectrum_1t} presents the SphinX spectra for the first and the second event respectively. In the top panel, there are observational spectra in counts per second without correction on the detector response. The low edge of the spectral measurements with SphinX is 1.2~keV which is typical for SphinX data and may be explained by the instrumental effects at low energies. In the bottom panel, the synthetic isothermal spectra are plotted. The spectra are shown measured at the moment of flare maximum, and in red are the preflare spectra considered as background. During the first event only the high-energy part of the emission changes significantly, making the spectrum flatter. Therefore this event is seen only in the short-wavelength part of the spectrum which is common for the majority of considered events. In the second event (the right panels in Figure~\ref{spectrum_1t}), the emission grows in both the low-energy and high-energy parts of the spectrum. Applying the two-temperature (2T) model allows an increase in the number of flares distinguished in the range 2.8\,--\,36.6~\AA. For about a half of the events in the 2T approximation, the 2.8\,--\,36.6~\AA\ flux during the flare maximum exceeds the preflare level. In any case, this spectral region is not good for flare detection and analysis and we use it mainly for study of active regions. Figure~\ref{profile} also demonstrates two main types of events from the viewpoint of their time profiles. The first one (left) is a short impulsive event, and the second one (right) is an event with a gradual decay which may be associated with mass ejections \citep{Kirichenko2013}. In this study we didn't distinguish these two types from one another and considered all of them as microflares.

\subsection{Processing of MISH Data}
   
MISH is a monochromatic soft X-ray spectroheliometer that takes images in the resonance doublet of the H-like ion Mg{\sc xii} ($\lambda$ 8.419 and 8.425~\AA) with a spatial resolution of about 2 arcsec. The appreciable emission in these lines forms only at a temperature of about 4MK and above. This feature strongly simplifies the process of hot coronal plasma localization on the solar disk. The disadvantage of MISH was the absence of coordinate information for its images. To solve this problem we used additional data from the EUV telescope FET onboard CORONAS-Photon. Our method was based on the coalignment of MISH images with close-in-time FET images using the known shift between FET and MISH fields of view. It is important that MISH is a calibrated instrument and allows us to obtain the luminosity of high-temperature plasma by simple integration of the emission source in the image.
\subsection{Processing of MDI and FET Data}
\begin{figure}    %%%%%%%%%%%%%%%%%% FIGURE 3 
   \centerline{\includegraphics[width=0.8\textwidth,clip=]{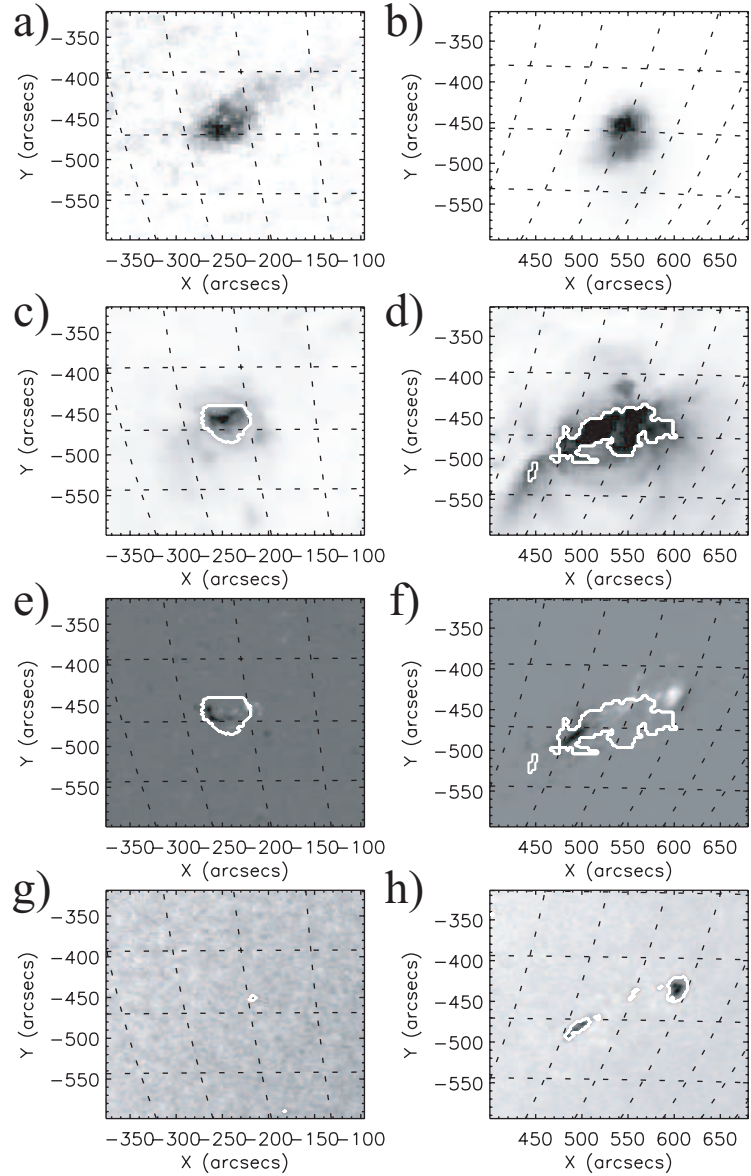}
              }
              \caption{Two microflares registered on 21 June 2009 with maximum at 20:10 (left images) and on 7 July with maximum at 23:45 (right images). Panels a-b: X-ray images of MISH; panels c-d: images of FET with flare region; panels e-f: MDI magnetograms with flare region contours; panels g-h: MDI intensitygrams and sunspot selection.
              }
   \label{source}
   \end{figure}

In order to calculate the strength of the magnetic field and its flux in the region of the microflare, we used Level 1.8 line-of-sight MDI magnetograms with a cadence of 96 minutes. The locations of the flaring regions were determined using data of MISH and FET. After that we co-aligned FET and MDI data to find these regions on the magnetograms. The period of observations, selected for the study, was characterized by a very low level of solar activity. Principally, during this period, we could observe only 1-2 small active regions on the disk simultaneously. Because of this we did not face any problems in the identification of active regions on the solar disk. The strength of the magnetic field was calculated in the following way (see Figure~\ref{source}). First, we selected a square frame of about 5$\times$5 arcmin size that contains the active region with the flare. From this frame, in the magnetogram we excluded pixels for which the unsigned values $B$ were $1\sigma$ lower than median unsigned $B$ through the frame. To select the flare location we used EUV images from FET telescope in the spectral line Fe{\sc xxxiii} 132~\AA. As a flare region, we considered the area of $3\sigma$ above the median emission level in the corresponding FET frame. To avoid overestimation of magnetic field values, we also excluded from mangetograms the sunspots seen on MDI intensitygrams. Then we averaged the unsigned magnetic data through the flare region to calculate the unsigned value $B$ for this flare. The magnetic flux was then calculated as $\Phi=B S$, where $ S $ is the area of the flare region. Figure~\ref{source} illustrates this procedure for two events (a weaker flare on the left side and a stronger one on the right).

  \begin{figure}    %%%%%%%%%%%%%%%%%% FIGURE 4 
   \centerline{\includegraphics[width=1\textwidth,clip=]{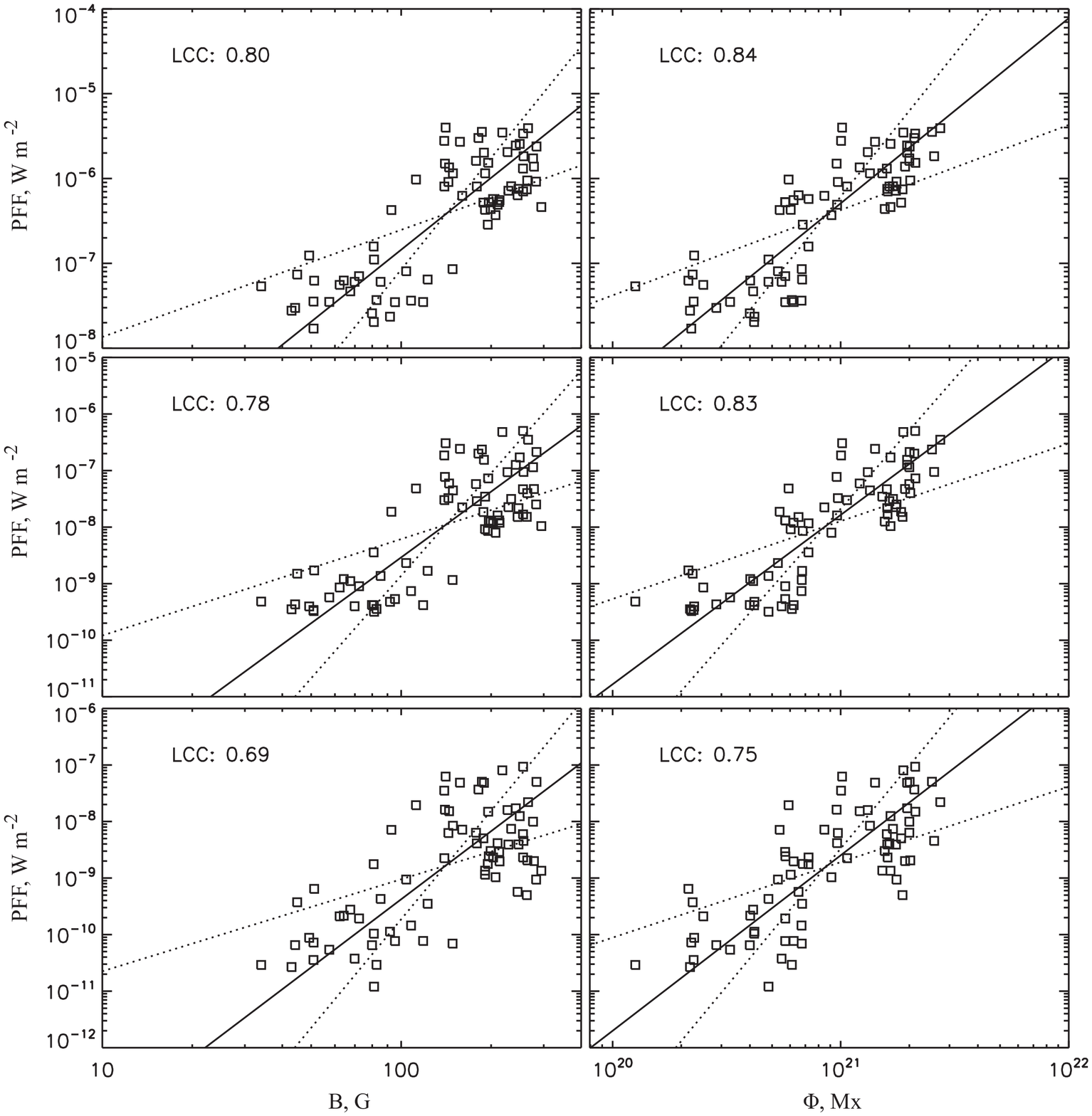}
              }
              \caption{ X-ray peak flare flux PFF \textit{vs.} average magnetic field strength $B$ (left plots) and \textit{vs.} total unsigned magnetic flux $\Phi$ (right plots). Top panel: results for the range 2.8\,--\,36.6~\AA. Middle panel: the same for the range 1\,--\,8~\AA\ for SphinX data. Bottom panel: results in 8.42~\AA\ for MISH data. In the 2.8\,--\,36.6~\AA\ range the emission of the active region dominates. In the 1\,--\,8~\AA\ range and at 8.42~\AA\ the results are mostly related to solar microflares.
                      }
   \label{all}
   \end{figure}
\section{Results and Discussion}
As stated above, we analyzed 163 solar microflares using TESIS and MDI data. The majority of events ($>$90\%) were registered in the latitude range 25\,--\,30$^{\circ}$, \textit{i.e.} in the zone of active regions located according to Sp{\"o}rer's law. To avoid projection effects we selected only events with a longitude within 50$^{\circ}$.
SphinX data allowed the analysis to be performed in any spectral range within an energy interval of 0.5\,--\,15 keV. The most interesting wavelength ranges for our study were 2.8\,--\,36.6~\AA\ and 1\,--\,8~\AA. The first one can be compared with data from \citet{Wolfson2000} and \citet{Pevtsov2003} \,--\, see Table~\ref{table1}. The range 1\,--\,8~\AA\ was used by \citet{Su2007}. The corresponding results for those intervals obtained at the flare maximum are presented in Figure~\ref{all}. The upper plots show data integrated over 2.8\,--\,36.6~\AA\ wavelengths. The middle plots were obtained for 1\,--\,8~\AA. In the bottom panel, we show the same distributions for monochromatic MISH data measured in the line 8.42~\AA. In our work, we use W m$^{-2}$ as the units of PFF. In general, all the relationships were approximated by power law functions using the least squares method. The solid lines on all plots correspond to the fit and the dotted lines show the 3$\sigma$ error. The accuracy of such a fitting can be characterized by linear correlation coefficients (LCC). All the results are summarized in Table~\ref{table_results}. As mentioned above, the emission in the spectral range 2.8\,--\,36.6~\AA\ mostly represents active regions while the emission in 1\,--\,8~\AA\ is mostly produced by the flare. The fitting results confirm this. Table~\ref{table_results} demonstrates that the indexes in the 1\,--\,8~\AA\ range and in the 8.42~\AA\ line are very close one to another, while the power-law indices in the range 2.8\,--\,36.6~\AA\ are significantly lower.

\begin{figure}[]  %%%%%%%%%%%%%%%%%% FIGURE 5 
   \centerline{\includegraphics[width=1\textwidth,clip=]{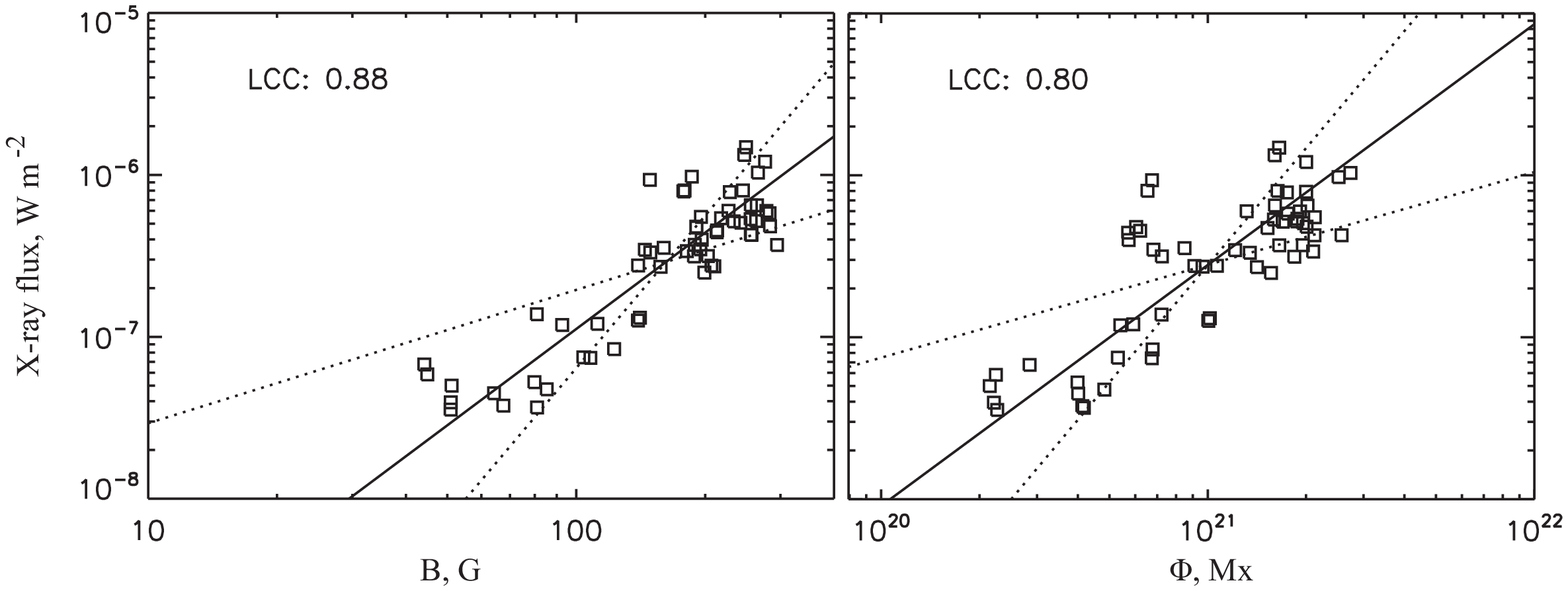}
              }
              \caption{X-ray flux in the range  2.8\,--\,36.6~\AA~in active regions \textit{vs.} average magnetic field strength $B$ (left plot) and \textit{vs.} total unsigned magnetic flux $\Phi$ (right plot).
                      }
   \label{active_soft}
   \end{figure}

\begin{figure}[]  %%%%%%%%%%%%%%%%%% FIGURE 6 
   \centerline{\includegraphics[width=1\textwidth,clip=]{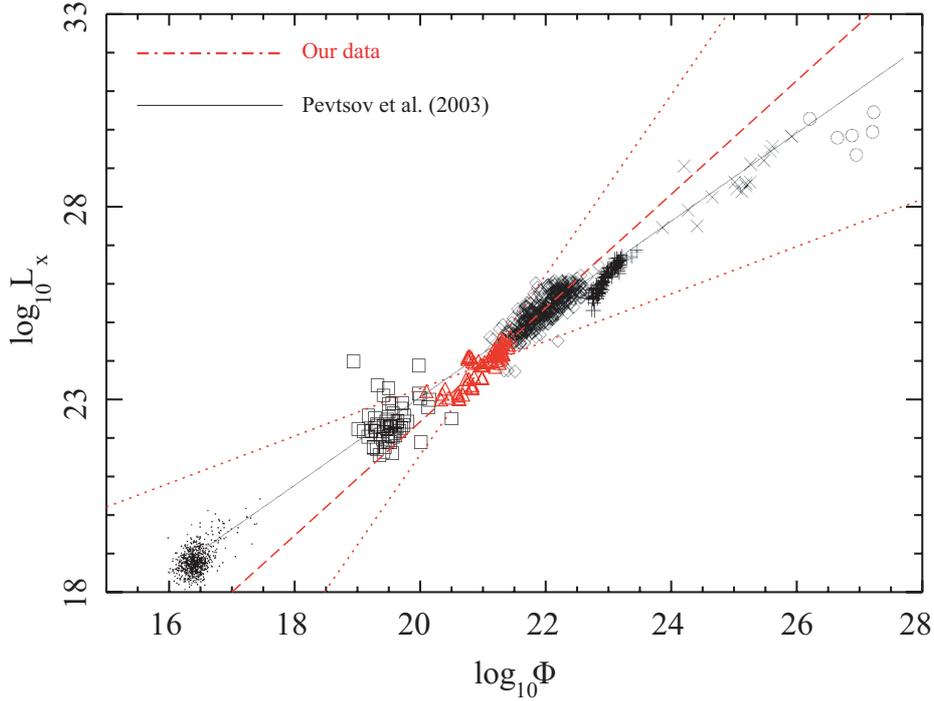}
              }
              \caption{Integrated X-ray luminosity in the range 2.8\,--\,36.6~\AA~\textit{vs.} total unsigned magnetic flux for solar and stellar objects. Dots: Quiet Sun. Square: X-ray bright points. Diamonds: Solar active regions. Pluses: Solar disk averages. Crosses: G, K, and M dwarfs. Circles: T Tauri stars. All data are from \citet{Pevtsov2003}. Triangles: Our data. Solid line: Power-law approximation for \citet{Pevtsov2003}. Dashed line: Power-law approximation for our data. Dotted lines: Error range for our data.
                      }
   \label{pevtsov}
   \end{figure}

The power-law index for the relation between X-ray flux in active regions ($F_{x}$) in the range 2.8\,--\,36.6~\AA and average magnetic field strength is $1.98\pm1.15$ (see Figure~\ref{active_soft}), which is close to the value 1.86 obtained by \citet{Wolfson2000} for SXT emission averaged along longitudes. As described in the introduction, \citet{Wolfson2000} found that their result is very close to the value 1.95 obtained by \citet{Roald2000} for the one-dimensional circularly symmetric supergranulation reconnection model. The index for $F_{x}(\Phi)$, namely $1.48\pm0.86$ is close to the result of \citet{Pevtsov2003}. To make the comparison with the data of \citet{Pevtsov2003}, we put our data and the results of \citet{Pevtsov2003} together in one figure (see Figure~\ref{pevtsov}). In order to unify the data, we transformed X-ray flux $F_{x}(\Phi)$ in W m$^{-2}$ to luminosity $L_{x}(\Phi)$ in erg s$^{-1}$ by the following way:
\begin{equation}\label{5}
L_{x}=4\pi R^{2}F_{x}\times10^{7},
\end{equation}
where $R$ is the Sun-Earth distance. Figure~\ref{pevtsov} demonstrates that the relationship $L_{x}$($ \Phi $) for our data in the range 2.8\,--\,36.6~\AA\ obeys, in general, the same law as the same relationship obtained for other solar and stellar magnetic structures in this spectral region. If we consider the different groups of solar objects from the data set of \citet{Pevtsov2003}, the best correspondence is achieved for active regions.

\begin{table}[b]
	\caption{Fitting results for PFF \textit{vs.} $B$ and PFF \textit{vs.} $\Phi$ flare distributions for different spectral ranges.}
   \label{table_results}	
	\begin{tabular}{ccccc}
	\hline
	Wavelength,~\AA & PFF($ B $) index & PFF($ B $) LCC & PFF($ \Phi $) index & PFF($ \Phi $) LCC \\
	\hline
	2.8\,--\,36.6 & 2.82 $\pm$ 1.56 & 0.8 & 2.18 $\pm$ 1.18 & 0.84 \\
	1\,--\,8  & 3.87 $\pm$ 2.16 & 0.78 & 3 $\pm$ 1.6 & 0.83 \\
	8.42  & 4 $\pm$ 2.37 & 0.69 & 3.1 $\pm$ 1.77 & 0.75 \\	
	\hline
	\end{tabular}
\end{table}

\begin{figure}[]  %%%%%%%%%%%%%%%%%% FIGURE 6
   \centerline{\includegraphics[width=1\textwidth,clip=]{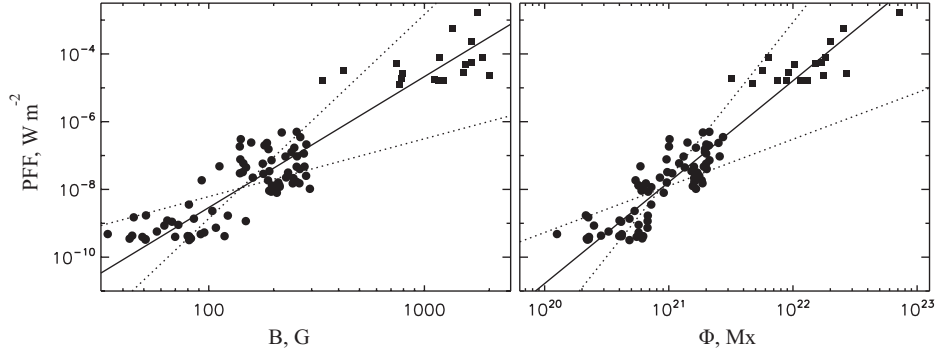}
              }
              \caption{Left: X-ray peak flare flux in the range 1\,--\,8~\AA\ \textit{vs.} the average magnetic field strength for our data (circles) and for data from \citet{Su2007} (squares). Right: X-ray peak flare flux in the range 1\,--\,8~\AA\ \textit{vs.} the magnetic field flux for our data (circles) and for data from \citet{Su2007} (squares).
                      }
                      
   \label{su}
   \end{figure}

In Figure~\ref{su}, we demonstrate PFF \textit{vs.} $B$ and PFF \textit{vs.} $\Phi$ distributions for solar flares in the range 1\,--\,8~\AA\ using our data for microflares (circles) and data from \citet{Su2007} for ordinary and strong flares (squares). In general, both distributions seem to be power-law functions with the power-law index of 3.87 for PFF$(B)$ and 3 for PFF$(\Phi)$ from our data and 0.93 and 1.11 from \citet{Su2007} respectively. We believe that such a huge difference between our results could be explained by the low statistics of the data in \citet{Su2007}. We don't try to fit them by a joint power-law equation for the following reasons. The first one is a possible difference in the proportion of total energy for the range 1\,--\,8~\AA\ between ordinary flares and microflares. Another reason is that the emission flux depends not only on the total but mainly on the non-potential part of the magnetic field energy. It means that for the same values of total magnetic flux the share of non-potential magnetic energy can be different. The absence of information on the non-potential energy of the magnetic field may lead to uncertainties in calculation of the power-law indices for the considered relations.

\begin{figure}    %%%%%%%%%%%%%%%%%% FIGURE 7 
   \centerline{\includegraphics[width=1\textwidth,clip=]{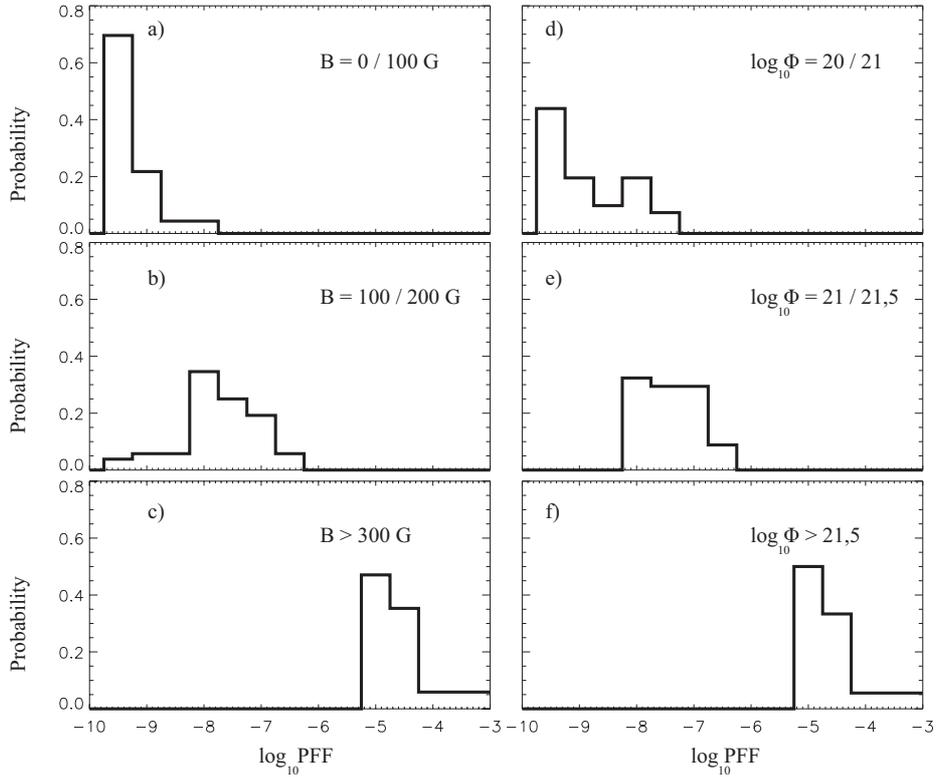}
              }
              \caption{Probability density functions to register a flare with a given PFF.
                      }
   \label{probability}
   \end{figure}

From the PFF \textit{vs.} $B$ and PFF \textit{vs.} $\Phi$ plots, it is possible to make some conclusions on the probability to detect a flare with a given X-ray class for different values of the magnetic field. The result is presented in Figure~\ref{probability}.

As described in the introduction, the majority of papers consider stable structures, such as nonflaring active regions, quiet corona, \textit{etc.} They use the hydrostatic solutions obtained in \citet{Rosner1978} under the assumption of constant pressure and heating rate. The assumption of constant pressure can be used in the case of loops which are shorter than a pressure scale height \citep{Aschwanden2004}. The hot and compact loops in microflares should also satisfy this condition.

  \begin{figure}    %%%%%%%%%%%%%%%%%% FIGURE 8 
   \centerline{\includegraphics[width=1\textwidth,clip=]{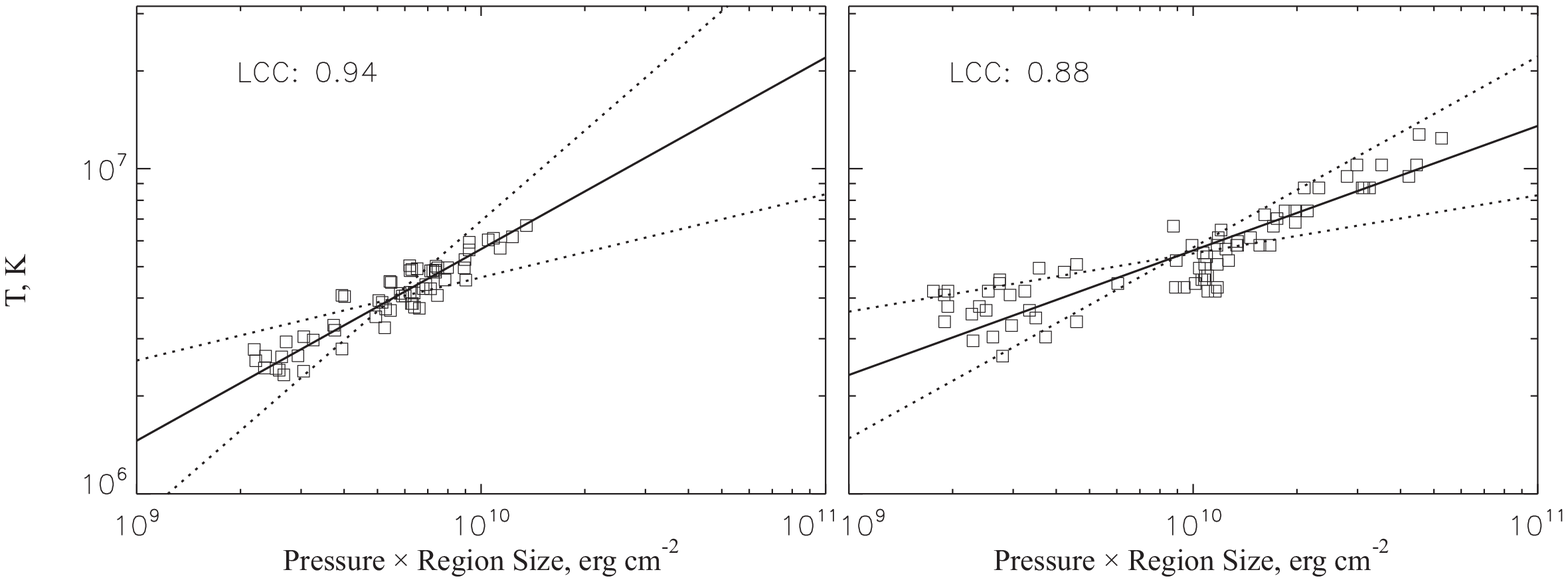}
              }
              \caption{Scaling law between the temperature and the combination of pressure and loop length for active regions (left plot) and microflares (right plot). Solid line \,--\, power-law approximation. Dotted lines \,--\, error range.
              }
   \label{t_pl}
   \end{figure}
 
Figure~\ref{t_pl} shows the correlation between temperature $T$ and the combination of pressure $p$ and loop length $l$ for active regions (left plot) and microflares (right plot). The temperature $T$ and emission measure EM for active regions were calculated by fitting the SphinX preflare spectra under the isothermal assumption. The same parameters for microflares were obtained by the 2T model at the maximum of microflares. We considered the temperature and emission measure of hot component as the flaring ones. The density $n_{e}$ and pressure $p$ were calculated as $n_{e}=($EM$/V)^{1/2}$, $p=2n_{e}k_{B}T_{e}$. The distance between the loop footpoints for microflares was estimated by  using the images of telescope FET in the line Fe{\sc xxiii} 132~\AA; then we calculated the volume assuming the loop is a semicircle. For active regions we obtained the area of the active region within the 3$\sigma$ contour and then calculated the volume assuming the structure as hemisphere.

The resulting relationships for active regions and microflares can be approximated by power-law functions with power-law indices $0.59\pm0.33$ and $0.42\pm0.2$ respectively. Both indices coincide with the RTV one (0.33) within the error range. Note that the index for active regions is slightly higher than for microflares.

   \begin{figure}    %%%%%%%%%%%%%%%%%% FIGURE 9 
   \centerline{\includegraphics[width=1\textwidth,clip=]{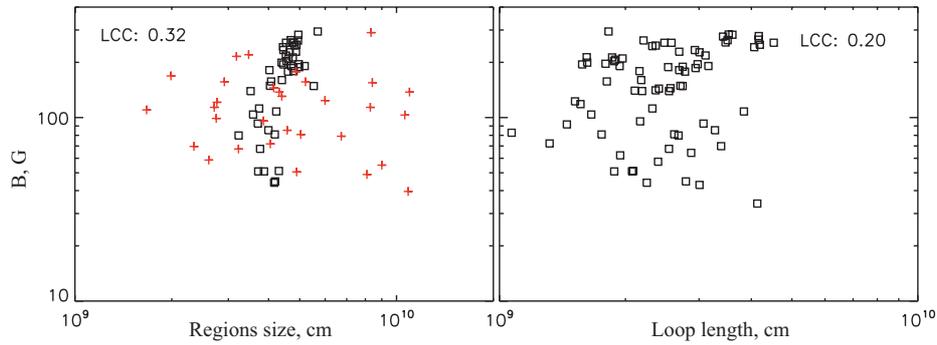}
              }
              \caption{Scaling law between the averaged magnetic field strength and region size for active regions (left plot) and for microflares (right plot). Squares: our data. Pluses: data of \citet{Yashiro2001}. 
              }
   \label{l_b}
   \end{figure}
   \begin{figure}    %%%%%%%%%%%%%%%%%% FIGURE 10 
   \centerline{\includegraphics[width=1\textwidth,clip=]{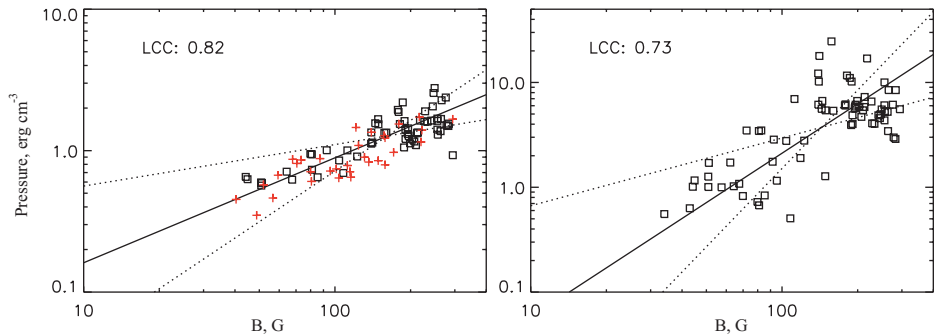}
              }
              \caption{Scaling law between the averaged magnetic field strength $B$ and pressure for active regions (left plot) and for microflares (right plot). Squares: our data. Pluses: data of \citet{Yashiro2001}.
              }
   \label{b_p}
   \end{figure}
   
As we mentioned above the coronal heating process can be described by Equations (\ref{1})--(\ref{4}). To obtain $\alpha$ or $\beta$ for Equation (\ref{1}) and Equation (\ref{4}) it is important to understand if there is any correlation between $B$ and $L$. In Figure~\ref{l_b} we demonstrate the relation between active region size (flaring loop size for microflares) and the magnetic field strength for active regions (left plot) and microflares (right plot). Unfortunately, we have a small number of unique active regions and we added the data from \citet{Yashiro2001} to our plot to increase the total statistics. The total LCC for active regions calculated for our data together with \citet{Yashiro2001} is 0.32 and the LCC for microflares is 0.20. Therefore we can consider the relation between $p$ and $B$ separately from the region size. Figure~\ref{b_p} demonstrates the correlation between $p$ and $B$ for active regions (left plot) and microflares (right plot). We also added the data of \citet{Yashiro2001} to the active region plot, but didn't use them for fitting. We just demonstrate the compatibility of our results. Using the least square method we obtained the power law indices $0.73\pm0.4$ and $1.53\pm0.8$ respectively. From Equation (\ref{4}) we have $\alpha=0.85\pm0.47$ for active regions and $\alpha=1.79\pm0.93$ for microflares.

Our results for microflares are very close to the results obtained by \citet{Golub1980} ($P\propto B^{1.6}$ with $\alpha=1.87$ from Equation (\ref{4})), while the data on active regions are close to the \citet{Yashiro2001} results ($p\propto B^{0.78\pm0.23}$ and $\alpha=0.91\pm0.27$). Thus, our results demonstrate that there should be two different heating mechanisms in pre-flaring active regions and during the microflares: Alfv\'en wave heating for active regions, and magnetic reconnection for microflares. The evidence supporting the presence of Alfv\'en wave turbulence heating in active regions has been presented by \citet{Fludra2017}.

\section{Conclusions}
In this paper, we examined relations between the parameters of X-ray emission and the characteristics of magnetic field for solar active regions and microflares of different X-ray classes, from A0.02 to B5.1. The results of the analysis demonstrate that for the microflares, as well as for large flares, there is a power-law relation between the X-ray peak flare flux (PFF) and the following magnetic field parameters: the total unsigned magnetic flux and average magnetic field strength.

In order to compare our results with the results of other authors, we calculated such relations for different spectral regions. The first one is 2.8\,--\,36.6~\AA\ where the emission is mostly produced by solar active regions. The contribution of solar flares here, as we found, is very small. The second and the third ones are the GOES range 1\,--\,8~\AA\ and the Mg{\sc xii} spectral line 8.42~\AA. These ranges are favorable to detect and study solar microflares. Our results for active regions in the range 2.8\,--\,36.6~\AA\ were found to be close to results obtained by \citet{Pevtsov2003} for different solar and stellar magnetic structures and \citet{Wolfson2000} for disk averages. The power law index for the relation $F_{x}$($ \Phi $) for active regions was found to be $1.48\pm0.86$ in our work and $1.13\pm0.05$ in \citet{Pevtsov2003}. For solar microflares we found the following relationships: power-law indices 3.87$\pm$2.16 and 3$\pm$1.6 for PFF($B$) and PFF($\Phi$) in the range 1\,--\,8~\AA, and 4$\pm$2.37 and 3.1$\pm$1.77 for PFF($B$) and PFF($\Phi$) in the line 8.42~\AA.

We made some suggestions on the heating mechanisms in active regions before and at the maximum of the microflares. For this purpose, we used RTV scaling laws under the assumption of constant pressure and heating. By analyzing the relation $P\propto B^{6\alpha/7}L$, we obtained $\alpha=0.85\pm0.47$ for active regions ($\alpha=1$ is the value predicted in the Alfv\'en wave heating scenario) and $\alpha=1.79\pm0.93$ ($\alpha=2$ is the value in the magnetic reconnection heating scenario). We believe that our results may be helpful for a deeper understanding of the physics of active regions and solar microflares and may help to answer the question of what mechanisms are responsible for heating in active regions, microflares and flares.

\begin{acks}
This work was supported by the Russian Science Foundation (RSF) grant № 17-12-01567. We also thank anonymous reviewer for careful revision of the work and very useful comments that significantly improved the manuscript.
\end{acks}

\begin{acks}[Disclosure of Potential Conflict of Interest]
The authors declare that they have no conflict of interest.
\end{acks}

\bibliographystyle{spr-mp-sola}
\bibliography{lit}  

\end{article}
\end{document}